\documentstyle[prd,aps,floats]{revtex}
\begin{document}
\draft

\input epsf \renewcommand{\topfraction}{0.8}
\twocolumn[\hsize\textwidth\columnwidth\hsize\csname
@twocolumnfalse\endcsname


\title{Challenges in generating density perturbations from
a fluctuating inflaton coupling}

\author{Kari Enqvist~$^{a,b}$, Anupam Mazumdar~$^{c}$,
and Marieke Postma~$^{d}$}
\address{$^{a}$ Department of Physical Sciences,
P. O. Box 64, FIN-00014, University of Helsinki, Finland.\\
$^{b}$ Helsinki Institute of Physics, P. O. Box 64, FIN-00014,
University of Helsinki, Finland.\\
$^{c}$ CHEP, McGill University, Montr\'eal, QC, H3A~2T8, Canada.\\
$^{d}$ The Abdus Salam International Centre for Theoretical Physics,
I-34100, Trieste, Italy.}
\maketitle

\begin{abstract}
We discuss the possibility of generating adiabatic density
perturbations from spatial fluctuations in the inflaton decay rate
which are due to quantum fluctuations of light moduli fields coupling
to the inflaton.  We point out that non-renormalizable operators,
which lift the flatness of the moduli potential, play an important
role for the density perturbations. In particular, the
non-renormalizable terms give rise to a considerable damping of the
fluctuations and thereby pose an obstruction to the construction of
possible models.
\end{abstract}

\vskip2pc]

\renewcommand{\thefootnote}{\arabic{footnote}}
\setcounter{footnote}{0}


Inflation is the main contender for an explanation of the observed
adiabatic density perturbations with a nearly scale invariant
spectrum~\cite{Linde}. However, recently alternative mechanisms for
generating the density perturbations have also been much discussed. In
the curvaton scenario, iscocurvature perturbations of some light
``curvaton'' field are converted into adiabatic perturbations in the
post-inflationary universe~\cite{Enqvist02,Enqvist03,Sloth}.  Another
interesting proposal is that the perturbations could be generated from
the fluctuations of the inflaton coupling to the Standard Model
degrees of freedom~\cite{Dvali,Kofman}. It has been argued that the
inflaton coupling strength to ordinary matter, instead of being a
constant, could depend on the vacuum expectation values (VEV) of the
various fields in the theory. If these fields are light during
inflation their quantum fluctuations will lead to spatial fluctuations
in the inflaton coupling strength. As a consequence, when the inflaton
decays, adiabatic density perturbations will be created because
fluctuations in the inflaton coupling translate into fluctuations in
the reheating temperature.

A particularly interesting implementation of this scenario is to
consider the Minimal Supersymmetric Standard Model (MSSM) plus an
inflaton field~\cite{Dvali}. There are many flat directions in the
MSSM~\cite{Enqvist}.  The moduli fields parameterizing these flat
directions are light, and their quantum fluctuations during inflation
produce fluctuations in the inflaton coupling.  The degeneracy of the
MSSM scalar potential is lifted by supersymmetry breaking effects and
by non-renormalizable operators.  In this note we analyze the effect
of these non-renormalizable operators on the produced density
perturbations.  We restrict the discussion to MSSM flat directions,
but the results can easily be adopted to more general models.

Assuming the inflaton is a gauge singlet, it can decay to normal
matter through both normalizable and non-renormalizable interactions
in the superpotential~\cite{Dvali}
\begin{equation}
W \ni \lambda_h \phi \bar{h} h +
\phi \frac{q}{M} q q + \phi \frac{q_c}{M} q_c q_c +
\phi \frac{h}{M} q q_c,
\end{equation}
where $\phi$ is the inflaton, $h$ and $\bar{h}$ are the two Higgs
doublets, and $q$ and $q_c$ are quark and lepton superfields and their
anti-particles. $M$ is some cut-off scale which could be the GUT scale
or the Planck scale.  The effective coupling for inflaton decay into
Higgs fields is $\lambda_h = \lambda_{0} (1+\frac{S}{M}+...)$, where
$S$ is the VEV of one of the flat direction fields in the theory. The
effective coupling for the inflaton decay to quarks is $\lambda_q =
\frac{S}{M}$ with $S= \langle q \rangle$, $ \langle q_c \rangle$, or
$\langle h \rangle$.  Effective couplings of this form can result from
integrating out heavy particles.

If a single decay channel dominates, the density contrast is
${\delta \rho}/{\rho} \sim \delta \lambda / \lambda$.\footnote{
It is also possible that one channel is responsible for the
fluctuations, but another is the main decay mode. Since the inflaton
decay is exponential $\propto \exp(-\Gamma/H)$ it would require some
fine-tuning to obtain sizable density perturbations.}
For $\lambda_q \gg \lambda_h$ the non-renormalizable couplings
dominate, and the inflaton decays predominantly into quarks and
anti-quarks. We will refer to this decay as indirect decay.  This
yields a density contrast
\begin{equation}
\label{pert0}
\frac{\delta\rho}{\rho}\sim \left.\frac{\delta S}{S}\right|_{\rm decay}\,.
\end{equation}
For light fields such that $m^2_S \ll H^2$ the quantum fluctuations
are set by the Hubble scale $\delta S \sim H$.  The density contrast
required to explain the observed temperature anisotropy in the cosmic
microwave background (CMB) radiation is $\delta\rho/\rho\sim 10^{-5}$,
which can be obtained for $S_* \sim 10^{5} H*$ provided there is no
later damping of the fluctuations. Here and in the following, the
subscript $*$ denotes the corresponding quantity evaluated at the time
observable scales leave the horizon, which is some 60 e-folds before
the end of inflation.  Since $S_* \gg H_{*}$ the perturbation spectrum
will be Gaussian.

In the opposite limit $\lambda_h \gg \lambda_q$ the inflaton decays
mainly into Higgs fields. We will call this the direct decay channel,
as decay is mediated by renormalizable operators.  The corresponding
density perturbation is
\begin{equation}
\label{pert00}
\frac{\delta \rho}{\rho}\sim \left.
\frac{\delta S}{M}\right|_{\rm decay}\,.
\end{equation}
For this decay channel the density perturbations are independent of
$S$. The fluctuations will be Gaussian for $S_* \gg H_{*}$ and
non-Gaussian in the opposite limit.  It is important to remember that
in Eqs.~(\ref{pert0},\ref{pert00}) the right hand side is always
evaluated at the time of inflaton decay.

As was pointed out in~\cite{Dvali} the flat direction $S$ must have a
mass smaller than $H_{*}$ in order to obtain adiabatic, nearly scale
invariant, fluctuations. Therefore the Hubble-induced supergravity
correction to the moduli mass should not exceed $m_{S}\sim
10^{-1}H_{*}$ in order to have a successful scenario. This can be
realized e.g. in no-scale supergravity models~\cite{Lahanas} with an
Heisenberg symmetry imposed on the chiral fields in K\"ahler
function~\cite{Gaillard}, or in models where inflation is driven by a
$D$-term.  Apart from the Hubble induced mass terms there are also
soft contributions from low energy supersymmetry breaking. However
such contributions are small $\sim {\cal O}(\rm TeV)$, and do not pose
any threat to the scenarios discussed in~\cite{Dvali,Kofman}. Finally,
the flat directions are lifted by non-renormalizable operators in the
superpotential of the form $W=\kappa{\cal
S}^{n}/nM_{p}^{n-3}$~\cite{Grisaru,Gherghetta}. Within the MSSM with
$R$-parity conservation most of the flat directions are lifted by
$n=4,5,6$ non-renormalizable operators~\cite{Gherghetta}. The flattest
one is lifted by $n=9$.  If dominant, these non-renormalizable
contributions can play an important role during and after
inflation.

Before discussing the effects of the non-renormalizable operators, let
us start by analyzing the parameter space where they are sub-dominant
and therefore can be neglected. The scalar potential for a flat
direction can be written as
\begin{equation}
\label{pot0}
V(S)=\frac{1}{2}m_{S}^2S^2
+\frac{\kappa^2 S^{2(n-1)}}{2^{n-1}M^{2(n-3)}}\,,
\end{equation}
where $\kappa \sim {\cal O}(1)$, and $m_{S}\sim {\cal O}(\rm TeV)$ is
the soft mass for the flat direction. We have assumed here that there
is no Hubble induced mass correction during inflation.  Such a mass
term can be included, but it will not change our conclusions in any
essential way. Requiring that the mass term dominates puts an upper
bound on the VEV during inflation
\begin{equation}
S_* \lesssim (m M^{n-3} / \kappa )^{1/(n-2)}.
\label{S}
\end{equation}

For the indirect decay channel $S_* \sim 10^5 H_*$, as follows from
Eq.~(\ref{pert0}).  This translates into an upper bound on the Hubble
constant during inflation: $H_* \lesssim 10^6,10^8,10^9 \, {\rm GeV}$
for respectively $n=4,\, 5, \,6$; here we have assumed $M \sim M_{\rm
pl}$, $\kappa \sim 1$ and $m_S \sim {\rm TeV}$. However, the bound can
be made stronger. Quantum fluctuations during inflation grow until a
saturation value $\langle S^2 \rangle \approx 3 H^4/8 \pi^2
m_S^2$~\cite{qf}. If one assumes that the VEV of $S$ has a typical
value $S_* \sim \sqrt{\langle S_*^2 \rangle}$, the bound on the Hubble
constant becomes $H_* \lesssim 10^3, 10^8, 10^{10}\, {\rm GeV}$ for
$n=4,\,5,\,6$ respectively.  This result is independent of $m_S$, except that $m_S
\sim 10^{-5} H_*$ in order to get density perturbations of the
observed size $\delta \rho/\rho \sim 10^{-5}$.  Taking $m_S \sim {\rm
TeV}$, this result is inconsistent for $n=4$, i.e., domination of the
mass term and density fluctuations of the observed size is
incompatible.  The results are only marginally consistent for $n=5$.

The density contrast generated through the direct decay channel is
independent of $S_*$, see Eq.~(\ref{pert00}).  A Gaussian perturbation
spectrum requires $H_* \lesssim S_*$, with $S_*$ bounded by
Eq.~(\ref{S}) if the mass term is to dominate.

We will now consider the opposite limit, in which the VEV of $S$ is
large and the non-renormalizable terms in the potential dominate.  We
should point out that the dynamics of the potential Eq.~(\ref{pot0})
has already been studied in \cite{Enqvist02,Enqvist03} in the context
of MSSM curvaton models. A simple analysis shows that during inflation
the flat direction field condensate is slow-rolling in the
non-renormalizable potential $V_{NR} \sim \kappa^2 S^{2(n-1)} /
M_{p}^{2(n-3)}$. The Hubble parameter and the amplitude of the field
can respectively be estimated as \cite{Enqvist03}
\begin{eqnarray}
\label{eqn1}
 H_* & \sim & \kappa^{-\frac{1}{n-3}}\delta^{n-2 \over n-3}M\,,\\
\label{eqn2}
 S_* & \sim & \kappa^{-\frac{1}{n-3}}\delta^{1 \over n-3}M \,,
\end{eqnarray}
where $\delta\equiv \delta S/S_* \sim H_*/S_*$.

Now let us turn our attention to the density perturbations.  The
equations of motion for the homogeneous and the fluctuation parts are
given by
\begin{eqnarray}
\label{pert1}
&&\ddot S+3H\dot S+V^{\prime}(S)=0\,,\\
\label{pert2}
&&\delta \ddot S_{k}+3H\delta\dot S_{k}+\left(\frac{k^2}{a^2}+V^{\prime\prime}
(S)\right)\delta S_{k}=0\,,
\end{eqnarray}
where prime denotes derivative w.r.t. $S$. Since we are only interested
in the long wavelength mode ($k\rightarrow 0$), using the slow roll
approximation during inflation we get
\begin{eqnarray}
\label{pert3}
&&3H\dot S+V^{\prime}(S)=0\,,\\
\label{pert4}
&&3H\delta \dot S+V^{\prime \prime}(S)\delta S=0\,.
\end{eqnarray}
where we have omitted the subscript $k$, understanding that $\delta S$
is for the superhorizon mode. The evolution of the ratio of the
fluctuations to the homogeneous mode of $S$ in the non-renormalizable
potential $V_{\rm NR}$ is
\begin{equation}
\frac{\delta S}{S}\sim \left(\frac{\delta S}{S}\right)_{\!{\rm i}}
\left(\frac{S}{S_{\rm i}}\right)^{2(n-2)}\,,
\label{pert5}
\end{equation}
where ${\rm i}$ denotes the initial value. During inflation the zero mode
obeys Eq.~(\ref{pert3}), which can be integrated to yield
\begin{equation}
\frac{S_{\rm end}}{S_{*}}\simeq
\left(1+\frac{1}{3(2n-3)}\frac{V^{\prime\prime}(S_{*})} {H^2}\Delta
N\right)^{-1/2(n-2)}\,,
\end{equation}
with $\Delta N$ the number of e-foldings after observable scales leave
the horizon, and $H_{\rm end}$ the Hubble constant at the end of
inflation. In a slow roll regime $V^{\prime \prime}(S_{*})\leq
H_*^2$. For $\Delta N = 60$ we find $S/S_{\rm i}\sim 0.7$ for
$n=4$. We conclude that during inflation the damping of the
perturbations is negligble small.

After inflation the moduli field $S$ slow-rolls (although marginally)
with $V^{\prime \prime}\sim H^2(t)$, and we can still use the slow
roll approximation Eqs.~(\ref{pert3},\ref{pert4}). The perturbations
are further damped according to Eq.~(\ref{pert5}) until $H \sim m_S$
and the field starts oscillating in the quadratic potential.  There is
no further damping during this epoch of oscillations.  The adiabatic
density perturbations are generated when the inflaton field decays,
which happens when $H \sim \Gamma_\phi \sim \lambda^2 m_\phi$, with
$m_\phi$ the mass of the inflaton.~\footnote
{We assume that the inflaton decays perturbatively.  The results
change if preheating occurs, and the inflaton decays through
non-perturbative processes.}
Thus, damping occurs between the end of inflation and, depending on
which event happens first, inflaton decay or the onset of moduli
oscillations.  The total damping factor is
\begin{eqnarray}
\label{fnl}
D \equiv \frac{ (\delta S/{S})_{\rm decay} }
{ (\delta S/{S})_{\rm end}}
&\sim&  \left(\frac{\max[\Gamma_\phi,m_S]}{H_{\rm end}}\right)^2 \nonumber \\
&\sim& \max \left[ \lambda_{\rm decay}^4, (m_S / m_\phi)^2 \right],
\end{eqnarray}
where we have used that at the end of slow roll inflation $H_{\rm
end}\simeq m_{\phi}$. For the indirect decay channel,
$(\lambda_q)_{\rm decay} = S_{\rm decay}/{M}$.  The correct level of
density perturbations is obtained if $D H_*/S_* \sim 10^{-5}$, as
follows from Eq.~(\ref{pert0}).  This implies the limit $S_{\rm decay}
\lesssim 0.1 M$ or $m_S \gtrsim 10^{-2} m_\phi$.  For the direct decay
channel on the other hand $(\lambda_h)_{\rm decay} = \lambda_0$ and
density perturbations are of the observed magnitude provided $D H_*/M
\sim 10^{-5}$, see Eq.~(\ref{pert00}).  Assuming $S_* \gtrsim H_*$ ,
which assures Gaussian fluctuations, this requires $\lambda_0 \gtrsim
0.1$ or $m_S \gtrsim 10^{-2} m_\phi$.

To conclude, the inclusion of non-renormalizable operators in the
potential of the moduli leads rather generically to a considerable
damping of the perturbations.  This puts severe constraints on the
parameters of the model.  In particular, either the inflaton coupling
to normal matter should be rather large, $\lambda \gtrsim 0.1$, or the
moduli mass should be large, $m_S \gtrsim 10^{-2} m_\phi$.  Another
possibility is to consider low scale inflation.  In this case the VEV
of the moduli is small, and the non-renormalizable operators in the
potential are subdominant.  However, for the indirect decay channel we
find that this appears to be inconsistent if the potential is lifted
by $n=4$ operators, and only marginally consistent for $n=5$. The
constraints are milder for the direct decay channel.

\section*{Acknowledgments}
A. M. is a Cita National fellow, K. E. is partially supported by the
Academy of Finland grant 51433, and M. P. is supported by the European
Union under the RTN contract HPRN-CT-2000-00152 Supersymmetry in the
Early Universe.
\vskip10pt




\begin{references}


\bibitem{Linde}
For a review, see: A.D. Linde, {\it Particle Physics And Inflationary
Cosmology}, Harwood (1990).

\bibitem{Enqvist02}
K.~Enqvist, S.~Kasuya and A.~Mazumdar,
Phys.\ Rev.\ Lett.\  {\bf 90}, 091302 (2003)

\bibitem{Enqvist03}
K.~Enqvist, A.~Jokinen, S.~Kasuya and A.~Mazumdar,
hep-ph/0303165.

\bibitem{Sloth}
K.~Enqvist and M.~S.~Sloth, Nucl.\ Phys.\ B {\bf 626}, 395 (2002)
D.~H.~Lyth and D.~Wands, Phys.\ Lett.\ B {\bf 524}, 5 (2002).
M.~Postma, Phys.\ Rev.\ D {\bf 67}, 063518 (2003).

\bibitem{Dvali}
G. Dvali, A. Gruzinov, and M. Zaldarriaga, astro-ph/0303591.

\bibitem{Kofman}
L. Kofman, astro-ph/0303614.

\bibitem{Enqvist}
For a review, see K.~Enqvist and A.~Mazumdar, hep-ph/0209244.

\bibitem{Lahanas}
For a review, see A. B. Lahanas and D. V. Nanopoulos,  Phys. Rept. {\bf 145},
1 (1987).

\bibitem{Gaillard}
M.~K.~Gaillard, H.~Murayama and K.~A.~Olive,
Phys.\ Lett.\ B {\bf 355}, 71 (1995).

\bibitem{Grisaru}
M. Grisaru, W. Sigl, and M. Rocek, Nucl. Phys. B {\bf 159}, 429
(1975); N. Seiberg, Phys. Lett. B {\bf 318}, 469 (1993);
M.~Dine, L.~Randall and S.~Thomas, Nucl.\ Phys.\ B {\bf 458}, 291 (1996).

\bibitem{Gherghetta}
T.~Gherghetta, C.~F.~Kolda and S.~P.~Martin,
Nucl.\ Phys.\ B {\bf 468}, 37 (1996).

\bibitem{qf}
T.~S.~Bunch and P.~C.~Davies,
Proc.\ Roy.\ Soc.\ Lond.\ A {\bf 360}, 117 (1978);
A.~D.~Linde,
Phys.\ Lett.\ B {\bf 116}, 335 (1982);
A.~A.~Starobinsky,
Phys.\ Lett.\ B {\bf 117}, 175 (1982);
A.~Vilenkin and L.~H.~Ford,
Phys.\ Rev.\ D {\bf 26}, 1231 (1982).


\end{references}
\end{document}